\documentclass[12pt]{article}

\def\unlockat{\catcode`\@=11}

\unlockat
\newcommand{\arstretch}[1]{\renewcommand{\arraystretch}{#1}}
\newcommand{\0}{{(0)}}
\newcommand{\1}{{(1)}}
\newcommand{\2}{{(2)}}
\newcommand{\3}{{(3)}}
\newcommand{\sinq}{\sin\theta\:}
\newcommand{\mon}{(\ref{eq:5}-\ref{eq:7})}
\newcommand{\domega}{d\Omega_2}
\newcommand{\bg}{\textrm{\scriptsize bg}}
\newcommand{\df}{\delta^3(\Sigma_2\hookrightarrow M_5)}
\newcommand{\jr}{\textsc{\scriptsize jr}}
\newcommand{\cs}{\textsc{\scriptsize cs}}
\newcommand{\inflow}{\mathrm{inflow}}
\newcommand{\fund}{\mathrm{\scriptstyle{fund}}}
\newcommand{\adj}{\mathrm{\scriptstyle{adj}}}
\newcommand{\zm}{\mathrm{\scriptstyle{z.m.}}}

\newcommand{\su}[1]{\ensuremath{SU(#1)}}
\newcommand{\so}[1]{\ensuremath{SO(#1)}}

\newcommand{\Dsl}{\mathop{\lefteqn{D}{\,\mbox{\large /}}}} 
\newcommand{\Asl}{\mathop{\lefteqn{A}{\,\mbox{\large /}}}} 

\newcommand{\R}{\ifmmode {2{\cal R}(0) - 1} \else $2{\cal R}(0) - 1$ \fi}
\newcommand{\Tr}{\mathop\mathrm{Tr}}      
\newcommand{\ch}{\mathop\mathrm{ch}}      
\newcommand{\U}{\ifmmode {\cal U} \else ${\cal U}$ \fi}
\newcommand{\V}{\ifmmode {\cal V} \else ${\cal V}$ \fi}

\newcommand{\brane}{{\ensuremath{\Sigma_2}}}
\newcommand{\bulk}{{\ensuremath{M_5}}}
\newcommand{\bundle}{{\ensuremath{S_\epsilon(\Sigma_2)}}}
\newcommand{\dfrac}[2]{\displaystyle\frac{#1}{#2}}
\newcommand{\dint}{\displaystyle\int}
\newcommand{\diag}{\mathop\mathrm{diag}}  
\newcommand{\eabc}{\epsilon_{abc}}
\newcommand{\g}[1]{\gamma^{#1}}

\newcommand{\st}{\ifmmode{M_5}\else ${M_5}$\fi}            
\newcommand{\sy}{(\sigma\cdot \y{})}               
\newcommand{\ws}{\ifmmode{\Sigma_2}\else ${\Sigma_2}$\fi}  
\newcommand{\y}{\hat y}

\def\IL{\relax{\rm I\kern-.18em L}}
\def\IH{\relax{\rm I\kern-.18em H}}
\def\IR{\relax{\rm I\kern-.18em R}}
\def\IC{\relax\hbox{$\inbar\kern-.3em{\rm C}$}}
\def\IZ{\relax\ifmmode\mathchoice
{\hbox{\cmss Z\kern-.4em Z}}{\hbox{\cmss Z\kern-.4em Z}}
{\lower.9pt\hbox{\cmsss Z\kern-.4em Z}}
{\lower1.2pt\hbox{\cmsss Z\kern-.4em Z}}\else{\cmss Z\kern-.4em
Z}\fi}

\def\CN {{\cal N}}

\def\CD {{\cal D}}

\def\CH {{\cal H}}

\def\CA{{\cal A}}

\def\CN {{\cal N}}

\font\manual=manfnt \def\dbend{\lower3.5pt\hbox{\manual\char127}}
 
\def\IZ{\relax\ifmmode\mathchoice {\hbox{\cmss Z\kern-.4em Z}}{\hbox{\cmss
      Z\kern-.4em Z}} {\lower.9pt\hbox{\cmsss Z\kern-.4em Z}}
  {\lower1.2pt\hbox{\cmsss Z\kern-.4em Z}}\else{\cmss Z\kern-.4em Z}\fi}
 \def\p{\partial}

\def\CN {{\cal N}}

\def\ch{{\rm ch}}

\def\IZ{\relax\ifmmode\mathchoice
{\hbox{\cmss Z\kern-.4em Z}}{\hbox{\cmss Z\kern-.4em Z}}
{\lower.9pt\hbox{\cmsss Z\kern-.4em Z}}
{\lower1.2pt\hbox{\cmsss Z\kern-.4em Z}}\else{\cmss Z\kern-.4em
Z}\fi}
\def\IA{\relax{\rm I\kern-.18em A}}
\def\IB{\relax{\rm I\kern-.18em B}}
\def\IC{{\relax\hbox{$\inbar\kern-.3em{\rm C}$}}}
\def\ID{\relax{\rm I\kern-.18em D}}
\def\IE{\relax{\rm I\kern-.18em E}}
\def\IF{\relax{\rm I\kern-.18em F}}
\def\IG{\relax\hbox{$\inbar\kern-.3em{\rm G}$}}
\def\IGa{\relax\hbox{${\rm I}\kern-.18em\Gamma$}}
\def\IH{\relax{\rm I\kern-.18em H}}
\def\II{\relax{\rm I\kern-.18em I}}
\def\IK{\relax{\rm I\kern-.18em K}}
\def\IP{\relax{\rm I\kern-.18em P}}

\def\inbar{\,\vrule height1.5ex width.4pt depth0pt}

\def\p{\partial}

\font\cmss=cmss10 \font\cmsss=cmss10 at 7pt
\def\IR{\relax{\rm I\kern-.18em R}}

\def\boxit#1{\vbox{\hrule\hbox{\vrule\kern8pt
\vbox{\hbox{\kern8pt}\hbox{\vbox{#1}}\hbox{\kern8pt}}
\kern8pt\vrule}\hrule}}
\def\mathboxit#1{\vbox{\hrule\hbox{\vrule\kern8pt\vbox{\kern8pt
\hbox{$\displaystyle #1$}\kern8pt}\kern8pt\vrule}\hrule}}

\def\inbar{\,\vrule height1.5ex width.4pt depth0pt}

\font\cmss=cmss10 \font\cmsss=cmss10 at 7pt
\def\IR{\relax{\rm I\kern-.18em R}}

\begin{document}
\title{\hfill\vbox{\normalsize\hbox{hep-th/0203154}
    \hbox{EFI-02-67}}
\\
     A Toy Model of the M5-brane: Anomalies of Monopole Strings
  in Five Dimensions.}
\author{ Alexey Boyarsky\footnote{boyarsky@alf.nbi.dk}~\footnote{On leave of
    absence from Bogolyubov ITP, Kiev, Ukraine}\\
  {\normalsize \it Niels Bohr Institute} \\
  {\normalsize \it Blegdamsvej 17, DK-2100 Copenhagen, Denmark}\\\\
  Jeffrey A. Harvey\footnote{harvey@theory.uchicago.edu}~~and
  Oleg Ruchayskiy\footnote{ruchay@flash.uchicago.edu} \\
  {\normalsize \it Enrico Fermi Institute and Dept. of Physics}\\
  {\normalsize \it University of Chicago} \\
  {\normalsize \it 5640 S. Ellis Ave. Chicago IL, 60637, U.S.A.}}
\date{}

\maketitle
\begin{abstract}
  We study a five-dimensional field theory which contains a
  monopole (string) solution with chiral fermion zero modes. 
  This monostring solution is a close analog of the fivebrane solution of
  M-theory. The
  cancellation of normal bundle anomalies parallels that for the M-theory
  fivebrane, in particular, the presence of a Chern-Simons term in
  the low-energy effective $U(1)$ gauge theory plays a central role. 
  We comment on the relationship between the the microscopic analysis
  of the world-volume theory and the low-energy analysis and draw
  some cautionary lessons for M-theory.

\end{abstract}

\section{Introduction}
\label{sec:anom-canc-m5}

The question of anomaly cancellation in the low-energy effective action of
M-theory in the presence of M5-brane was discussed in~\cite{witten} where
non-cancellation of the normal bundle anomaly was first found. This issue was
addressed later by several authors from different points of view (see
e.g.~\cite{Bonora,fhmm,Lechner} and discussions therein). In this paper we are
continuing to follow the path of~\cite{fhmm}.

The spectrum of a theory containing solitonic brane solutions can be decomposed
into normalizable modes living on the brane, and non-normalizable bulk
modes. 
If one approximates these as decoupled theories, then one can find
inconsistencies at the very beginning. In particular, if some of the zero
modes are chiral and the dimension of the world-volume is even (as for the
M5-brane or solitonic string case), the world volume theory suffers from
gravitational and gauge anomalies. Thus it cannot be considered by itself and the
two theories -- bulk and world-volume  -- interact in such a way that the
whole theory is anomaly free. In general one finds that
the bulk action has a gauge variation which is non-zero only on the brane, and
this variation cancels the gauge variation arising from the localized chiral
zero modes. Such cancellation of anomalies is sometimes
referred to as the ``inflow mechanism''~\cite{ch}. 

As far as gravitational anomalies are concerned, the world-volume zero modes  transform
under local Lorentz transformations with the structure groups of tangent and
normal bundles of the brane's word-volume. The latter can be treated as a
gauge symmetry from the world-volume theory point of view. Anomalies of the
chiral fields of this theory are in general derived from anomaly polynomials,
given by Atiyah-Singer index densities.  This expression is not symmetric with
respect to normal and tangent bundles.  On the other hand, the naive variation
of gravitational Chern-Simons terms in the bulk effective action is symmetric
with respect to normal and tangent bundles.  Therefore some other contribution
to gravitational anomaly inflow from the bulk is required.

The work~\cite{fhmm} was based on the idea that for the M5-brane the
additional contribution comes from the $C \wedge d C \wedge d C$ Chern-Simons
term of eleven-dimensional supergravity when properly defined in the presence
of an M5-brane.  The Chern-Simons is \emph{defined} in terms of descent of a
globally defined gauge invariant closed form on a manifold of one dimension
higher.  In the presence of an M5-brane, the differential of the naive
Chern-Simons is no longer gauge-invariant.  Therefore we should construct a
globally defined, closed, gauge invariant form in one dimension higher first.
This was done in~\cite{fhmm} regularizing the M5-brane source, solving the
regularized magnetic coupling equation (Bianchi identity) and constructing a
closed form out of this solution.  The Chern-Simons term defined by such a
procedure also gets regularized -- it changes near the brane -- and this
regularization is not invariant under normal bundle gauge transformations.
This gives an additional contribution to the normal bundle anomaly inflow 
only and cancels the anomaly.

The price paid for this anomaly canceling mechanism was the presence in the
action of a \emph{bump function}: an arbitrary function of the distance from
the brane with definite boundary values.  This treatment left open the
question of the physical origin of the bump function $\rho(r)$ and a
microscopic derivation of the modified Chern-Simons term.  The origin of the
bump function was further addressed in a simple field theory model
in~\cite{jh-or}.


In other examples of anomaly cancellation, one can often understand the
cancellation in terms of the anomaly freedom of an underlying short distance
theory. In the example of axion strings~\cite{jh-or,ch}, there is a low-energy
effective theory containing chiral zero modes on the axion string, and a
coupling of the axion to $F \wedge F$ which cancels the anomaly in the
presence of a topologically non-trivial string. In the underlying theory, the
fermion measure is gauge invariant, and the anomaly cancellation reflects the
decomposition of this measure into a localized zero mode part and the remaining
massive, bulk degrees of freedom. The axion-gauge coupling arises from
integrating out the massive fermion degrees of freedom and clearly must cancel
the zero mode anomaly since the full fermion measure is gauge invariant.

In M-theory, no such corresponding microscopic picture that would explain the
observed anomaly cancellation is known. It would thus be interesting to know
if there are other models where cancellation of normal bundle anomalies arises
from a similar modification of a Chern-Simons term, and where this term can be
derived from a consistent underlying theory.


Another interesting question raised by anomaly cancellation using Chern-Simons
terms is the question of the scaling of the anomaly of the world-volume theory
with the magnetic charge $N$ of the brane~\cite{hmm}. The Chern-Simons term is
cubic in the gauge field which, naively speaking, should mean that the anomaly
inflow coming from the Chern-Simons term and, therefore, the anomaly of the
world-volume theory, should scale like $N^3$.  The M5-brane world volume
theory, the so called (2,0) theory, is very special and there are independent
arguments for an $N^3$ dependence of related quantities
\cite{klebs,gubser,henns}.  All these arguments (including the inflow argument
itself) are indirect. It is very interesting to see another example of the same
kind, with anomaly inflow coming from cubic Chern-Simons terms, but where the
theory of the zero modes is known explicitly and the corresponding anomaly can be
calculated directly.

To analyze these and other questions raised in~\cite{fhmm}, and to try to
get some possible insight on the high-energy structure of M-theory, we are
going to consider a simple field theory model, which however turns out to be
very similar to the original M-theory setup.  The theory in hand
(Section~\ref{sec:setup}) is just the theory of the 'tHooft-Polyakov monopole
lifted to 5-dimensional space-time and coupled to fermions.  The monopole
solution, viewed as a solution in five dimensions which is static and
independent of the fifth coordinate $x_4$, looks like a magnetically charged
string (monostring).  We analyze the theory of zero modes on the string and
find that it is anomalous (Section~\ref{sec:Jackiw-Rebbi}).  We show that
gravitational and gauge Chern-Simons terms do appear in the low energy
effective action of the bulk theory although no $SU(2)$ Chern-Simons in
five dimensions is allowed (Section~\ref{sec:induced-cs}).  We discuss
the correct definition of the Chern-Simons term on in the monostring background and show
that it varies under normal bundle transformations even in the limit of zero
monopole core i.e.  without regularization as in~\cite{fhmm,bk} (which
solves the above stated problem of appearance of an arbitrary bump function in
the effective action\footnote{A similar construction for the M5-brane case has been
recently presented in \cite{bk}}).  This gives the possibility to have a
non-symmetric inflow of anomalies with respect to normal and tangent bundles
which indeed cancels the anomalies of the zero modes.  This situation closely
resembles the M5-brane, yet is not precisely the same, because in our case we
know the full underlying \su2 theory and not just an effective description. We
also discuss the anomaly analysis in the low-energy effective theory
(Section~\ref{sec:other-anomalies}) and discuss the relation to the picture in
M-theory. Apart from that we analyze the situation of monopole solution with
charge $N$ (Section~\ref{sec:big-charge}), and show that in this case there is
no argument for an $N^3$ dependence of the anomaly inflow (despite the presence
of a  cubic Chern-Simons term!), in contrast to the situation in M-theory.

\section{The Model}

\label{sec:setup}

We start by considering a five-dimensional theory with \su2 gauge group, Higgs
fields in the adjoint representation, and fermions in either the fundamental
or adjoint representation (later we briefly consider fermions in arbitrary
representations of \su2). This theory is not renormalizable, but we will only
be concerned with the anomaly structure for which renormalizability is
irrelevant. We take the Lagrangian to be\footnote{see Appendix~\ref{sec:dyon}
  for conventions} (in the absence of gravity):
\begin{equation}
  \label{eq:1}
  {\cal L} = -\frac 14 (\Tr G_{MN})^2 + \frac 12 (D_M\Phi)^2  - \frac 1
  {g^2} U(g\Phi) + i\bar\psi_n \Dsl \psi_n - Gg\bar\psi_nT^a_{nm}\psi_m \Phi_a
\end{equation}
Here $M,N = 0,\dots,4$ label coordinates in the bulk; indices $i,j,k = 1,2,3$
will be used later for bulk indices transverse to the string solution;
$m,n=1,\dots,\dim(\mathbf{r})$ are flavor indices and run up to the dimension
of the fermion representation \textbf{r}; $a = 1,2,3$ label generators of
$su(2)$ with commutation relations $\left [ T^a ,\, T^b \right ] =
i\epsilon^{abc}T^c$. We take the potential $U(\Phi)$ to be
\begin{equation}
  \label{eq:2}
      U(g\Phi) = \dfrac {\lambda^2v^2 g^2}{2} \left ( 1 - \frac 1 {v^2}
        \Phi^a\Phi_a\right )^2 
\end{equation}
In the fundamental representation we can choose
\begin{equation}
  { T^3 = \diag (1/2,-1/2) }
  \label{eq:3}
\end{equation}
while in the adjoint representation we take
\begin{equation}
{ T^3 = \diag (1,0,-1)}\label{eq:4}
\end{equation}
This theory can also be coupled to gravity in the obvious way.
%
%
%

The Lagrangian~(\ref{eq:1}) admits a monopole solution~\cite{monopole} which
is independent of $x^0,x^4$ and varies only in the $y^i$ directions: 
%
\begin{eqnarray}
  \label{eq:5}
   && A_0^a = 0\\
   && A_i^a = -\epsilon_{aij} \hat y^j A(y)/g\label{eq:6}\\
   && \Phi^a = \hat y^a \phi(y)/g \label{eq:7}
\end{eqnarray}
Both $\phi(y)$ and $A(y)$ vanish at $y=0$; for large $y$ $\phi(y) \rightarrow
g v$ and $A(y) \rightarrow -1/y$ exponentially.  In $4 + 1$ dimensions this
solution, supplemented with the condition $A_4^a = 0$, describes a magnetically
charged string, stretched along the $x^4$ direction. Note that the solution
~\mon\ is invariant under a diagonal $SO(3)_{\rm diag}$ transformation with
generators $\vec K = \vec J + \vec T$  with $\vec J$ the generators of spatial
rotations and $\vec T$ the generators of $SU(2)$ gauge transformations. This
symmetry will play an important role in the later analysis.

We want to consider the theory~(\ref{eq:1}) on the monopole background~\mon\ 
and explore the close analogy between anomaly cancellation in this theory with
that of the M5-brane theory~\cite{fhmm}. We will see that by integrating out massive
fermions in this background we will obtain a theory closely resembling
$D=11$ SUGRA.

\subsection{Abelian Gauge}
\label{sec:abelian-gauge}

Throughout this paper we will use the \emph{Abelian gauge} condition ~(see
e.g.~\cite{monopole,freund}).  By performing the gauge rotation $ G =
\exp(-i\varphi T^3)\exp(i\theta T^2)\exp(i\varphi T^3)$ (where $\theta,\phi$
are spherical angles in transverse dimensions, $G\in\su2$) we rotate the field
$\Phi^a$ given by~(\ref{eq:7}) into $\tilde\Phi^a = \phi(y)\delta^{a3}$. Such
a transformation (which effectively can be implemented only \emph{outside the
  core} i.e. when the profile of the gauge field $\phi(y)$ is not zero) is of
course singular (it changes the topological charge of the field $\Phi$
configuration from 1 to zero):
\begin{equation}
  \label{eq:8}
  Q_{\mathrm{top}}= \frac 1 {4\pi} \int_{S^2_\infty} \eabc \Phi^a d\Phi^b
  \wedge d \Phi^c
\end{equation}
This gauge transformation also rotates the  gauge field~(\ref{eq:6}) into the
following configuration (outside the core): $\tilde A^{1,2}_i$ is equal to
zero and $\tilde A_i^3$ is given by the gauge potential for
a Dirac monopole (with the Dirac string along say the positive $y^3$ semi-axis):
\begin{equation}
  \label{eq:9}
  \tilde A_i^3 = -\frac 1g \epsilon_{i3k}\frac {\y^{k}}{|y|-y^3}
\end{equation}
Following 'tHooft~\cite{monopole} we  introduce the  $U(1)$ field strength
\begin{equation}
  \label{eq:10}
  \begin{array}{rl}
    F_{M N} &= \hat \Phi^a G^a_{M N} - \dfrac 1g \epsilon_{a b c} \hat
  \Phi^a D_M\hat \Phi^b D_M\hat \Phi^c \\
  & = \p_M(\hat \Phi^a A^a_N)- \p_N(\Phi^a A^a_M) + \dfrac 1g \epsilon_{a b c}
  \hat \Phi^a \p_M\hat \Phi^b \p_N\hat \Phi^c
\end{array}
\end{equation}
where $\hat\Phi^a = \Phi^a/v$. This object is \su2 invariant, defined outside
the core, and obeys the anomalous Bianchi identity:
\begin{equation}
  \label{eq:11}
  d F = \frac{4\pi}{g}\df,
\end{equation}
showing that there is a magnetic monopole source for $F$. 
For brevity we are using the language of differential forms
in~(\ref{eq:11}). The precise meaning of the source term  $\df$ will be explained later
(Section~\ref{sec:in-core}).

\section{Jackiw-Rebbi zero modes}
\label{sec:Jackiw-Rebbi}
 
Jackiw and Rebbi~\cite{Jackiw-Rebbi} considered the Hamiltonian for the Dirac
operator (in $3+1$ dimensions) in the monopole
background~\mon. It has the following form:
\begin{equation}
  \label{eq:12}
  \hat {\cal H}_D \psi = \left [ \vec\alpha\cdot  \vec p_y\, \delta_{nm} + A(y)
    T^a_{nm}( \vec\alpha \times \y)_a + G\phi(y) T^a_{nm} \y^a \beta\right
  ] \psi_m = E\psi_n 
\end{equation}
Here $\vec\alpha=\g0\vec\gamma$, with the  gamma-matrix conventions given in
Appendix~\ref{sec:gamma-matrices}.  This equation has \emph{zero modes} --
solutions with $E=0$ for fermions in both the fundamental and adjoint
representations.  The Jackiw-Rebbi zero mode for both representations has the
following form:
\begin{equation}
  \label{eq:13}
  \psi^\jr_n = {\chi^+_n \choose \chi^-_n}
\end{equation}
where $\chi^{\pm}_n$ are Weyl spinors.
The zero energy solutions for both representations have $\chi^-_n=0$. For
fundamental fermions we have:
\begin{equation}
  \label{eq:14}
  \chi^+_{\alpha n} = c\,\sigma^2_{\alpha n} f(y)
\end{equation}
Here $n = 1,2$ is a  color index, $\alpha = 1,2$ is a spinor index, and $c$ is
a normalization constant. The function $f(y)$ is given by (we denote $|y|$ by $y$)
\begin{equation}
  \label{eq:15}
  f(y) = \exp\left(-\int^y_0 dy'\Bigl(\frac 12 G \phi(y') - A(y')\Bigr)\right)
\end{equation}

For the  adjoint representation the zero modes  are given by
(see~\cite{Jackiw-Rebbi} for details):
\begin{equation}
  \label{eq:16}
  \chi_n^+ 
  = N\biggl[ \tilde f_1(y)\y^n \sy +  f_2(y)\sigma^n \biggr] s 
\end{equation}
with $N$ a  normalization constant and  $s$ an  arbitrary two-component spinor on
which the  $\sigma$-matrices act. We won't need the explicit 
functions $f_{1,2}(y)$, see \cite{Jackiw-Rebbi}, eqs.~(3.8)-(3.10)), the
only important property being that they depend on $|y|$ and not on $\y$. Also,
for convenience we have introduced a  new function (compared with $f_1,f_2$
from~\cite{Jackiw-Rebbi}): $\tilde f_1(y) \equiv f_1(y) - f_2(y)$.

Note, that the existence and the number of these zero modes is dictated by the
Callias index theorem~\cite{Callias}.

We note the following: if~(\ref{eq:14}),~(\ref{eq:16}) are zero modes of
the equation~(\ref{eq:12}) (which we can think of as the Dirac equation in 3
dimensions), then we can use them to build a solution of
the Dirac equation in $4+1$ dimensions in the
background~\mon. This is done in the following way.

Consider the  fundamental case first. Construct the following spinor out of
the solution~(\ref{eq:14}): 
\begin{equation}
  \label{eq:17}
  \Psi_n(x,y) = c(x)\psi^\jr_n(y)
\end{equation}
Here $\Psi$ is a five-dimensional spinor, $c(x)$ an  arbitrary (complex) scalar
function, and  $\psi^\jr_n(y)$ is given by~(\ref{eq:13}-\ref{eq:14}). The fermionic
e.o.m.'s, following from~(\ref{eq:1}) can be written as
\begin{equation}
  \label{eq:18}
  \Dsl\nolimits_{4+1}\Psi = i\g0(\p_0 + \g{int}\p_4 - i \hat\CH_D)\Psi = 0
\end{equation}
The definition of $\g{int}$ as the  chirality matrix in the two-dimensions of
the string world-sheet is
given  by~(\ref{eq:67}). Now substitute the ansatz~(\ref{eq:17})
into~(\ref{eq:18}). Note once again that both spinor and color indices are
carried by $\psi^\jr$. Also note that $\g{int}\psi^\jr= -\psi^\jr$. As a
result,the  equation $\Dsl\Psi=0$ boils down to
\begin{equation}
  \label{eq:19}
  (\p_0 -\p_4)c(x) = 0
\end{equation}
We see that our model possesses a  chiral (left-moving) zero mode, localized on
the string and falling off exponentially (as $e^{-v y}$ at large $|y|$) in the
transverse dimensions.

In a totally similar manner one can build the  zero modes of~(\ref{eq:18}) 
for adjoint fermions. The difference being that in this case there are two zero
modes~(\ref{eq:16}). The ansatz has the following form then:
\begin{equation}
  \label{eq:20}
  \Psi_n(x,y)=\sum_{A=1}^2 s_A(x)\psi^\jr_{n,\,A}
\end{equation}
The index $\textsc{a}$ is
a spinor index of the transverse \so3. Substituting the ansatz~(\ref{eq:20}) into
eq.~(\ref{eq:18}) we will get an  answer similar to~(\ref{eq:19}):
\begin{equation}
  \label{eq:21}
  (\p_0-\p_4)s_A = 0\quad{\textsc{a}=1,2}
\end{equation}
We will be also interested in finding the  action for these zero modes.
Substitute~(\ref{eq:17}--\ref{eq:20}) into the action for fermions:
\begin{equation}
  \label{eq:22}
  S_D=\int d^5x\, \bar\Psi(i\Dsl - G g\Phi)\Psi
\end{equation}
Here we consider $\Dsl$ in the background of the monopole gauge field plus
some (small) fluctuations in the $(x^0,x^4)$ directions, which we call
$a^a_\mu$ (recall that indices $\mu,\nu$ run along the world-sheet directions
of the string).

First, consider  fundamental fermions. Then eq.~(\ref{eq:22}) can
be written as:
\begin{equation}
  \label{eq:23}
  \begin{array}{rcl}
    S_D[\Psi] &= &\int d^5x \,i\Psi^+ (D_0 + \g{int}D_4 + \alpha^i D_i + i G g
    \g0\Phi)\Psi \\
    & = & \int d^2x \Bigl(\int d^3y f^2(y)\bar\sigma^2_{\alpha
      n}\sigma^2_{\alpha m}\Bigr) 
    i c^*(x)(D_0-D_4)_{nm}c(x) 
  \end{array}
\end{equation}
(we have $\chi^+_n\chi_m = f^2(y)\sigma^2_{\alpha n}\sigma^2_{\alpha m} =
f^2(y)\delta_{nm}$). This means that eq.~(\ref{eq:23}) is equivalent to:
\begin{equation}
  \label{eq:24}
   S_D[c] = N\int d^2x \,i c^* (\p_0-\p_4)c
\end{equation}
Here $N=2\int d^3 y\,f^2(y)$ is a finite normalization constant. The gauge field
$a_\mu^a$ has dropped out of $(\Dsl_{0,4})_{nm}$ because of $\delta_{nm}$. We
see that the  1+1 dimensional fermions effectively couple to  $\Tr a^a_\mu$,
which is zero for any \su2 connection. We see that fundamental fermions are
singlets with respect to the diagonal $SO(3)_{\rm diag}$ defined
earlier  (c.f.\cite{Jackiw-Rebbi}) This fact
can be checked directly: acting on the fermion~(\ref{eq:14}) with the
transformation of $\so3_{\diag}$ leaves it invariant.

For adjoint fermions  we substitute the  ansatz~(\ref{eq:20}) into the
action~~(\ref{eq:22}). This gives:
\begin{equation}
  \label{eq:25}
    S_D[\Psi] = \int d^2x \,\sum_{A,B=1}^2is^*_B (\Dsl)_{nm}s_A \int
    d^3y\chi^+_{n,B}\chi_{m,A} 
\end{equation}
we note that $\sum_n\chi^+_{n,B}\chi_{n,A} \sim \delta_{AB}$ and that the indices
$\textsc{a,b}$ are acted upon by $\sigma$-matrices from~(\ref{eq:16}). As a
result one obtains:
\begin{equation}
  \label{eq:26}
  S_D[s] = \int d^2x \,c_1 s^*_A(\p_0-\p_4) s_A +s^*_A
  \Asl{}^a\sigma_{AB}^a s_B 
\end{equation}
here $c_1 = \int d^3y\, (f_1^2 + f_2^2)$ and the ``gauge field'' on the brane is
given by the  expression:
\begin{equation}
  \label{eq:27}
  A^a_\mu = \int d^3y\,2i\tilde f_1f_2 (\y^b a^b_\mu)\y^a + 2if_1f_2a^a_\mu
\end{equation}
From~(\ref{eq:26})-(\ref{eq:27}) we see that in this case the zero modes become
charged with respect to $SO(3)_{\rm diag}$  and transform as (iso)spinor
representation of \so3  (c.f.~\cite{Jackiw-Rebbi}).
The original spinor indices of $\sigma_{AB}$ become gauge indices of \su2. This is
another exhibition that the zero modes provide a  representation of the diagonal 
$SO(3)_{\rm diag}$.


Theories of chiral fermions in two dimensions suffer from gauge and
gravitational anomalies. The magnetic string solution decomposes the tangent bundle to the
five-dimensional space-time $M_5$ in the usual way as $T M_5|_\Sigma = T\Sigma
\oplus N$ and correspondingly the symmetry group of the tangent bundle \so{4,1}
breaks into $\so{1,1}\times\so3$. The group \so3, the structure group of the normal
bundle to the string, becomes a gauge group from the point of view of
two-dimensional fermions (the same as $\so3_{\mathrm{diag}}$).  Using the
standard descent formalism \cite{as,fads,zum,anomrev} we can write down
anomalies associated with the fermion zero modes on the string.  We find that
the tangent bundle anomaly on string world-sheet plus $SO(3)_{\rm diag}$ anomaly with
respect to the group of normal bundle $N$ is given by
\begin{equation}
  \label{eq:28}
  \arstretch 2
  \begin{array}{lcl}
    I_\zm^\fund & = & \dfrac {\pi} {12}\int_\brane p_1^\1(T\Sigma)\cr
    I_\zm^\adj & = & \displaystyle\int_\brane\left(\dfrac\pi{6}
      p_1^\1(T\Sigma)-\dfrac\pi 2 p_1^\1(N)\right) 
  \end{array}
\end{equation}
Note that the anomalies are non-symmetric between tangent and normal bundles.  On
the other hand, the corresponding terms in the bulk effective action, which
are due to interaction with gravity, are expected to depend on the full $4+1$
dimensional spin connection because local Lorentz symmetry is not broken in
the bulk.  Thus the contribution to inflow from such terms will by symmetric with
respect to tangent and normal bundles as it was for the  M5-brane~\cite{witten}
case (as we will see below, this is also true in our model).This means that
some other sources of anomalies should be found.  This problem was originally
resolved in the context of an  analysis of the M5 brane~\cite{fhmm} by showing that the
Chern-Simons term of $D=11$ SUGRA has an additional anomalous variation.  At first
sight such a resolution cannot be applied in our case because no Chern-Simons
term is allowed in $4+1$ dimensions for an  \su2 gauge theory.  Any such term
would arise from descent of $\Tr F^3$ but this vanishes in $\su2$ (since there
is no $d^{abc}$ coefficient in $SU(2)$). In Section~\ref{sec:parity} we will
see what actually happens.

\section{Effective Action in the Bulk}
\label{sec:bulk}

In this section we will implement the procedure, described in the introduction,
to obtain a low-energy effective action.
As mentioned earlier, the monopole solution breaks the  \su2 gauge 
symmetry down to $U(1)$ and makes
the fermions massive, with masses given by $G v$
(c.f.~(\ref{eq:30}--\ref{eq:31})).  We consider the low-energy effective
action 
after integrating out these massive (charged) fermions. This action will be a function of
the $U(1)$ gauge field and of the gravitational field if we couple the theory
to gravity. We will assume we have done this in the standard way.

Integrating out the massive fermions exactly is a difficult task, but we can
go to Abelian gauge (Section~\ref{sec:abelian-gauge}) first and it will reduce
the theory down to a $U(1)$ gauge theory coupled to massive charged fermions as described
below (Section~\ref{sec:parity}), at least outside the core of the monopole.
All massive modes vanish near the core of monopole, and thus  far from defect the
theory looks topologically trivial (in Abelian gauge). Therefore we believe
that in integrating over massive fermions we can use the standard (perturbative)
approach (as in~\cite{Redlich,Reuter}). In addition, we will only be
interested in the parity non-invariant contributions to the effective action
which play a role in anomaly cancellation.

\subsection{Parity Violation}
\label{sec:parity}

As is usual in odd-dimensions, a parity transformation changes the sign of an
odd number of spatial directions and the fermion mass term changes sign under
this transformation (see e.g. the Appendix of~\cite{ahw}). Thus, (taking into
account that $U(\Phi)$ is an even function of $\Phi$), the above action has a
parity symmetry given by
\begin{equation}
x_4 \rightarrow -x_4,\qquad  \psi \rightarrow \g4 \psi, 
\qquad \Phi^a \rightarrow - \Phi^a\label{eq:29}
\end{equation}
The monopole solution~\mon\ not only breaks $\su2$ down to $U(1)$, but it also
breaks the parity symmetry. However the product of $P$ and charge conjugation
$C$ (which takes $A_M \rightarrow - A_M$ with $A_M$ the $U(1)$ gauge
potential and interchanges positive and negative charged states) remains a
symmetry. For example, after symmetry breaking $\su2 \rightarrow U(1)$ we
have a mass term for fundamental fermions $\chi_\pm$ with $U(1)$ charge $\pm
g/2$ of the form
\begin{equation}
  \label{eq:30}
  G v ( \bar \chi_+ \chi_+ - \bar \chi_- \chi_-) 
\end{equation}
while for adjoint fermions $\psi_+,\psi_0,\psi_-$
with charges $g,0,-g$ we have
\begin{equation}
  \label{eq:31}
  G v ( \bar \psi_+ \psi_+ - \bar \psi_- \psi_-).
\end{equation}
In either case we have two fermions with opposite charge and opposite mass.
These mass terms are invariant under parity plus the interchange of $\chi_\pm$
or $\psi_\pm$.  For adjoint fermions one field  will remain massless and
uncharged and will be present in the massless spectrum of the  bulk effective
theory.

Note that 
after symmetry breaking a Chern-Simons term for the $U(1)$ gauge field $\int A
\wedge F \wedge F$ {\it is}  allowed, and is consistent with the unbroken $CP$
symmetry. This does not contradict  the comment made at the end of
Section~\ref{sec:Jackiw-Rebbi}. The Chern-Simons term cannot be made out
of the \su2 connection alone, but must also involve the scalar Higgs fields.  
We will see that the Chern-Simons term  and other terms are
induced at the one-loop level by integrating out the massive fermions.

\subsection{Induced Chern-Simons Term}
\label{sec:induced-cs}

Following the results of~\cite{Redlich,Reuter,agdpm}, integrating out 
the  massive
fermions will induce, among other terms,  parity odd Chern-Simons (CS) terms in the
effective action. A fermion of mass $m$ in a representation ${\bf r}$
generates the C-S term
%
\begin{equation}
S_\cs=- \pi{m \over |m|} \int Q_5 \label{eq:32}
\end{equation}
where $dQ_5$ is given by the six-form part of the Dirac index
density
\begin{equation}
d Q_5 = \hat A(R) \ch_r(F)|_{6}\label{eq:33}
\end{equation}

We now apply this to our model with fermions of charge $\pm q$ with
$q = g/2,g$ for fundamental or adjoint fermions and with opposite
sign masses  to obtain
\begin{equation}
S_\cs = \pi \int Q_5^- - Q_5^+\label{eq:34}
\end{equation}
where
\begin{equation}
  \arstretch 2
  \begin{array}{rl}
    d Q_5^\pm & = \hat A(R) \ch_\pm(F)|_6 \\
    & = \displaystyle(1-{1 \over 24}p_1 + \cdots)(1 \pm {q \over 2 \pi} F + {1 \over 2!}
    ({ q \over 2 \pi})^2 F^2 \pm {1 \over 3!} ({ q \over 2 \pi})^3 F^3 +
    \cdots)\label{eq:35}
  \end{array}
\end{equation}
which gives 
\begin{equation}
    S_\cs = \int_\bulk-{ q^3 \over 24 \pi^2} A \wedge F \wedge F + { q \over 24}
F \wedge p_1^\0(R)\label{eq:36}
\end{equation}
(In this expression we take the $U(1)$ gauge field to be Hermitian.)

Note the close analogy between this action and the action for M
theory~\cite{dlm,witten}. The latter contains a CS term $\int C_3 \wedge G_4
\wedge G_4$ as well as the term $\int G_4 \wedge X_8^\0$ with $X_8$ being an
eight-form in curvature.  We might also expect in analogy to M theory, that
the $F \wedge p_1(R)^\0$ term is involved in cancellation of both tangent and
normal bundle anomalies of the world-sheet zero modes while the $A \wedge F
\wedge F$ term will play a role in the cancellation of normal bundle
anomaly. We will see that this is indeed the case. 

\section{Anomaly cancellation for $N=1$}
\label{sec:anom-canc}

There are three sources of anomalies: the fermion zero modes, inflow from the
$F \wedge p_1^\0(R)$ coupling, and a contribution to the normal bundle anomaly
from the CS term as in~\cite{fhmm}. In the following we discuss the structure
of the CS terms and the cancellation of anomalies in this model.

\subsection{Discussion of the  Chern-Simons terms}
\label{sec:in-core}

Following~\cite{fhmm,bk} we want to  correctly define the Chern-Simons term in
the bulk effective action.  It was implied in eqs.~(\ref{eq:33}--\ref{eq:36})
that $d F=0$ and so (at least locally) $F = dA$. We would like to consider
gauge fields in the same topological class as the background monopole solution
$F=F_\bg+d\tilde A=d(A_\bg+\tilde A)$, where $A_\bg$ could be defined only
locally.  To specify it explicitly, we take a closer look at~(\ref{eq:10}).
On the background~\mon\ it has the form (again in the language of differential
forms):
\begin{equation}
  \label{eq:37}
  F = F_\bg = \frac 1g\epsilon_{abc} \y^a d\y^b \wedge d\y^c
\end{equation}
(from now on we will put $g=1$).  One can recognize the right hand side
of this expression as being the  volume form on the two-sphere:
$\Omega_2:\quad\int_{S^2}\Omega_2 = 4\pi$.  We generalize it for the case of
non-trivial gravitational field and curved string world-sheet volume form as
$F_\bg = \frac {2\pi}{g} e_2 $.  Here by $e_2$ is the  \emph{global angular
  form}:
\begin{equation}
  \label{eq:38}
  e_2 = \frac 1{4\pi} \epsilon_{abc} \left(d \hat y^a d \hat y^b \hat y^c -
  d(\Theta^{ab} \hat y^c)\right) = \frac 1{4\pi} \epsilon_{abc} (D \hat y^a \land
  D\hat y^b y^c - R^{ab}(\Theta)\hat y^c) 
\end{equation}
where  $\Theta^{ab}$ is an \so3 connection on the normal bundle and
$R(\Theta)$ is the  corresponding curvature (see~\cite{BottTu} for rigorous
definitions or Appendix in~\cite{beckers} for a list of useful formulae and
properties). Note also the striking resemblance of the expression~(\ref{eq:38})
to  't Hooft's definition of the  Abelian field strength~(\ref{eq:10})!  For
$F_\bg = 2\pi e_2$ the one-form $A$ would be given by
\begin{equation}
A_\bg = {2\pi} e_1^\0\label{eq:39}
\end{equation}
Obviously, the form $e_1^\0$ is not globally defined.
 
Now, let us repeat formulae~(\ref{eq:33}--\ref{eq:36}) substituting explicitly
$F=F_\bg+d\tilde A$ and $A=A_\bg+\tilde A$ (we use $\tilde A$ to denote only
the globally well  defined part of the gauge potential).  Then ~(\ref{eq:36}) will be
modified to:
\begin{equation}
  \label{eq:40}
  \arstretch{2}
  \begin{array}{rl}
  S_\cs  = &\displaystyle\int_\bulk-\dfrac{ q^3}{24 \pi^2}({2\pi} e_1^\0+ \tilde A) \wedge (2\pi
  e_2+d\tilde  A) \wedge(2\pi e_2+d\tilde  A)\\
  & + \dfrac{ q}{24}  (2\pi e_2+d\tilde  A)\wedge p_1^\0(R)
\end{array}
\end{equation}

Let us recall that the one form $e_1^\0$ is not gauge invariant under $SO(3)$
rotations of normal bundle.  Therefore part of the Chern-Simons term is not
invariant as well. Its gauge variation will be calculated in the
Section~\ref{sec:anom-canc-1} and we will see that it gives additional normal
bundle anomaly inflow we were looking for.  The reason for this seems to be
the following -- with local definition of $A_\bg$, Chern-Simons is usually
defined patch by patch. When we do $SO(3)$ rotation of the normal bundle, we
mix patches and Chern-Simons needs to be transformed. To see this effect we
need some topologically non-trivial field configuration which in case at hand
is represented by fluctuations around monopole (monostring)
background~(\ref{eq:37}).  Technically this effect is reflected in the fact
that gauge potential of any field from this topological class contains the
form $e_1^\0$ which has non-zero $SO(3)$ variation.

All this was done in the bulk, outside the  monopole core where the Abelian theory is
defined. We see that for the Chern-Simons term  it is not necessary to be
regularized to give normal bundle anomaly inflow -- it just a property of
topologically non-trivial background. The same procedure can be done also for
the M5-brane case~\cite{bk}.

Let us see now what happens if we want to regularize the Chern-Simons term
nevertheless
and compare with~\cite{fhmm}.  There the approach was the following. First, to
have forms defined everywhere we should regularize the magnetic source
$\df$. The natural way to do it is to consider a
\emph{Poincare dual}~\cite{BottTu} of the submanifold $\Sigma$. It is given~by
\begin{equation}
  \label{eq:41}
  \tau(\Sigma) = \frac 12 d\rho\wedge e_k
\end{equation}
where $e_k$ is global angular $k$-form, and the monotonic function $\rho(|y|)$
equals  $-1$ on the submanifold, then goes to zero fast enough and is zero
outside tubular neighborhood of the submanifold $\Sigma$. The equation of
magnetic coupling~(\ref{eq:11}) (or the modified Bianchi identity) then has the
form:
\begin{equation}
  \label{eq:42}
  d F = 2\pi d\rho\wedge e_k
\end{equation}
The above formulae are the same for any magnetic source of even codimension.
The only difference between M5 and our case is that $k=4$ in the former case
while here $k=2$ and $F=dC_3$ instead of $F=dA_1$.  Now we have to find the
solution of magnetic coupling equation~(\ref{eq:41}) defined everywhere.
The general solution is given by $F=Ae_k+B\rho e_k+Cd\rho e_{k-1}+(\mbox{exact
  form})$, where $A,B,C$ are arbitrary constants.  In~\cite{fhmm} the solution
with $A=0,B=0,C=1$ was chosen. To compare this with our case let's recall that
we already have the regularized version of our string -- this is the
't Hooft-Polyakov monopole (monostring) of the  5-d $SU(2)$ theory. Indeed, let's
consider~(\ref{eq:10}) evaluated on the monopole background~\mon :
\begin{equation}
  \label{eq:43}
  F = \frac1{ v^3}\epsilon_{abc} \Phi^a d\Phi^b \wedge d\Phi^c =  2\pi\frac
  {\phi^3(y)}{v^3} e_2 
\end{equation}  
Comparing this with the expression~(\ref{eq:41}) for the Poincare dual we find
the relation:
\begin{equation}
  \label{eq:44}
  1+\rho(y) = \frac{\phi^3(y)}{v^3}
\end{equation} 
As a result we come to the conclusion that in the monopole background, the 
(regularized) Abelian field strength is given by:
\begin{equation}
  \label{eq:45}
  F = 2\pi(1+\rho)\,e_2
\end{equation}
This corresponds to a solution of the generalized Bianchi identity with
$A=1,B=1,C=0$. Let us take this solution and proceed further along the lines
of~\cite{fhmm}. We have to construct out of this solution a closed form.
Exactly as in~\cite{fhmm} and by the same reason we can choose it to be
\begin{equation}
  \label{eq:46}
  d A = F - 2\pi\rho\,e_k
\end{equation}
(Again, as it was above, the form $A$ here is defined only locally because
$dA$ does not need to be (and, in fact, is not) exact, but only closed). If we
substitute our choice for the $F$ we see that $dA=2\pi e_k+\mbox{(exact
  form)}$ and $A=2\pi e_{k-1}^\0+\mbox{(globally~defined~form)}$.  As a result the
``regularized'' Chern-Simons term (obtained by the same analysis as in~\cite{fhmm}
but applied to another solution of magnetic coupling equation) is the same as
it was above~(\ref{eq:40}), before regularization, and does not contain
function~$\rho(y)$ (as opposed to the case~\cite{fhmm})\footnote{This last
part of the analysis is also exactly the same for the M5 case~\cite{bk}.}.  This
seems to be a solution of one of the problems formulated in the Introduction.

\subsection{Anomaly Cancellation I}
\label{sec:anom-canc-1}

Let us now calculate the anomaly inflow from the bulk and then put all the
different contributions to the anomaly together.

First consider
the term $d A \wedge p_1^\0(TM)$:
\begin{equation}
  \label{eq:47}
  \delta_{TM}\left[ \int_\bulk d A \wedge p_1^\0(TM)\right] = \int_\bulk  d A \wedge d
  p_1^\1(TM) = \int_\bulk d\left (  d A \wedge p_1^\1\right)
\end{equation}
Recall that the integration in~(\ref{eq:47}) goes along $M_5$ \emph{minus} the tubular
neighborhood of the string world-sheet $D_\epsilon(\Sigma)$. Thus the integration
in eq.~(\ref{eq:47}) reduces to that  on the boundary of $\bundle = \p
D_\epsilon(\Sigma)$:
\begin{equation}
  \label{eq:48}
  -\int_\bundle d A \wedge p_1^\1 = -4\pi\int_\brane p_1^\1(TM) =
  -4\pi\int_\brane \left(p_1^\1(T\Sigma) \oplus p_1^\1(N)\right)
\end{equation}
where we used the explicit form of $A$ (eq.~(\ref{eq:39})) and the fact that
integral of $e_2$ along the fiber is equal to $2$. The minus sign in
eq.~(\ref{eq:48}) comes from the fact that the positive orientation of the
\bundle\ is with respect to the brane \brane\ and not to the bulk. As already
noted at the end of Section~\ref{sec:Jackiw-Rebbi} this inflow is symmetric
between tangent and normal bundle to the string and so no values of the
coefficient would cancel the anomaly~(\ref{eq:28}) of the zero modes.

Now we consider the $A\wedge dA \wedge dA$ term. It has a non-zero
variation  only with
respect to the \so3 normal bundle due to $\delta_N e_1^\0 = d e_0^\1$.
Following~\cite{bott-c,fhmm} we write:
\begin{equation}
  \label{eq:49}
  \arstretch 2
  \begin{array}{rl}
    \delta_N \left[ \int_\bulk A(dA)^2\right] &\equiv (2\pi)^3 \int_\bulk d
  e_0^\1e_2^2 \\
  & = -(2\pi)^3\int_\bundle e_0^\1e_2^2 = - 2(2\pi)^3\int_\brane  p_1^\1(N)
\end{array}
\end{equation}
Bringing the results of~(\ref{eq:48}),~(\ref{eq:49}) together
with~(\ref{eq:40}) we can write:
\begin{equation}
  \label{eq:50}
  \arstretch 2
  \begin{array}{lcl}
    I_\cs & = &\dfrac{2 \pi q^3}{3}\dint_\brane p_1^\1(N) \cr
    I_\inflow & = &-\dfrac{q \pi}{6}\dint_\brane \left(\vphantom{\dfrac{q
          \pi}{6}} 
      p_1^\1(T\Sigma) + p_1^\1(N) \right) 
  \end{array}
\end{equation}
where $q=1/2,1$ for fundamental or adjoint fermions. Remarkably, the anomalies
cancel for both fundamental and adjoint fermions~(\ref{eq:28}).

We stress once again that the bump function $\rho(y)$ does not enter directly
in the anomaly cancellation procedure. The anomalous variation of the
Chern-Simons term is simply a consequence of evaluating it on the
topologically nontrivial monopole background ($e_2$ in this case). This
evaluation can always be done outside the core and will not involve the
function~$\rho$.  We see, that the additional normal bundle anomaly inflow
comes not because of microscopic modifications inside the core, but rather
due to the fact that we have to carefully define the Chern-Simons in the
presence of the topologically non-trivial configuration.

\section{Anomaly Cancellation II\\ Arbitrary Fermion Representation}
\label{sec:arbitr-rep}

To make the story even more convincing we would like to demonstrate that this
anomaly cancellation actually works for arbitrary fermion representation.
Consider fermions with the arbitrary isospin $T$. To compute the tangent
bundle anomaly it suffices to know the total number of zero modes which is
given by the index theorem of Callias~\cite{Callias}.  It tells us that the
total number of zero modes for the fermions in the arbitrary representation
$T$ is:
\begin{equation}
  \label{eq:51}
  \CN_\zm = \left\{
    \begin{array}[c]{ll}
      T(T+1)&\quad \mbox{$T$ -- integer}\\
      T(T+1)-\dfrac14&\quad \mbox{$T$ -- half-integer}
    \end{array}
  \right.
\end{equation}
%
However, to determine the normal bundle anomaly we also need to know
how these zero modes transform under $\so3_{\diag{}}$. As far as we
are aware, this information is not available in the literature, and
thus we cannot provide a definitive test of anomaly cancellation.
We will however show that a simple and natural guess for the structure
of the fermion zero modes does lead to complete anomaly cancellation.
It would be interesting to try to verify this guess by explicit
computations. 

Any wave-function $\psi_{\alpha\,n}(x)$ realizes the tensor product of two
representations: rotational \so3 with a spin $J$ (where $J=L+1/2$ is the total
angular momentum) and \su2 gauge symmetry with isospin $T$.  The tensor product
$J\otimes T$ can be decomposed in terms of \so3 representations as
$|J-T|\oplus\dots\oplus|J+T|$. To find what representation of $\so3_{\diag{}}$
the  zero modes realize, we choose the following strategy: adjust the
angular momentum $L$ of the zero modes 
so as to achieve the lowest possible total spin of in the
tensor product. It is clear from~(\ref{eq:51})
that the  zero modes will then transform in a  reducible representation of \so3.

\paragraph{Integer Isospin:}
\label{sec:integer-isospin}
Consider the case of integer $T$ first. The wave-function of zero modes transforms as
$J\otimes T$ and we choose $J=T+1/2$. Then
\begin{equation}
  \label{eq:52}
  (T+1/2)\otimes T = 1/2\oplus 3/2\oplus\dots\oplus(2T+1/2)
\end{equation}
One can easily check that the total dimension of the first $T$ terms
in~(\ref{eq:52}) (up to the spin $(T-1/2)$) is equal to $T(T+1)$, as we need
in view of~(\ref{eq:51}). Let us check that the anomaly indeed cancels against
the bulk. 

The CS term in the action is just the sum of those similar to~(\ref{eq:40})
for $q=1,2,\dots,T$. As a result, anomaly inflow is given by the sum of analogs
of~(\ref{eq:50}) with different $q$'s:
\begin{equation}
  \label{eq:53}
  \arstretch 2
  \begin{array}{lcl}
    I_\cs & = &\dfrac\pi{6}T^2(T+1)^2 \dint_\brane p_1^\1(N) \cr
    I_\inflow & = &-\dfrac{\pi}{12}T(T+1)\dint_\brane \left(\vphantom{\dfrac{q \pi}{6}}
      p_1^\1(T\Sigma) + p_1^\1(N) \right) 
  \end{array}
\end{equation}
The anomaly of the zero modes is not difficult to find using 
the usual descent
formalism. 
\begin{equation}
  \label{eq:54}
  I_\zm =-2\pi\int_\brane\Bigl[\hat
  A(R)\:\ch_\mathbf{r}(F)\Bigr|_{\mbox{\raisebox{-5pt}{$\scriptstyle 4$}}}\Bigr]^\1    
\end{equation}
where the Chern character $\ch_\mathbf{r}(F) = \Tr_\mathbf{r}\exp\frac{F}{2\pi}$.
And the Chern character for arbitrary representation is 
%
$\mathbf{r}$ with the spin $\frac{2t-1}{2}$ is:
\begin{equation}
  \label{eq:55}
   \ch_\mathbf{\frac{2t-1}{2}}(F) = 2t + \frac{t(4t^2-1)}{12}p_1(N) + \cdots
\end{equation}
The total anomaly is the sum of anomalies for $t=1,\dots,T$, and is equal to:
\begin{equation}
  \label{eq:56}
  I_\zm^T = \int_\brane\left(\frac{\pi}{12}T(T+1)\,p_1^\1(T) - \frac{\pi}{12}
    T(T+1)(2T^2+2T-1)\,p_1^\1(N)\right) 
\end{equation}
We see that this expression cancels~(\ref{eq:53}).

\paragraph{Half-integer Isospin:}
\label{sec:half-integer-isospin}
This case is very similar to  the previous one. The total number of zero modes is given
by~(\ref{eq:51}), and if we write $T=n+1/2$, where $n$ is a natural number,
then $\CN_\zm = n^2$. We choose that $J=T$ in this case (again, to be able to
get the lowest spin with respect to $\so3_{\diag{}}$ in $J\otimes T$). 
As a result the  zero modes form the reducible representation with the
$n$ lowest spins
\begin{equation}
  \label{eq:57}
  T\otimes T = \underbrace{0\oplus 1\oplus\dots\oplus(n-1)}_{\mbox{\small $n$ terms}}
    \oplus\dots\oplus 2n -1
\end{equation}
We repeat the procedure described for the integer isospin.  The inflow terms give
us:
\begin{equation}
  \label{eq:58}
  \arstretch 2
  \begin{array}{lcl}
    I_\cs & = &\dfrac\pi{12}n^2(2n^2-1)\dint_\brane p_1^\1(N) \cr
    I_\inflow & = &-\dfrac{\pi}{12}n^2\dint_\brane\left(\vphantom{\dfrac{q \pi}{6}}
      p_1^\1(T\Sigma) + p_1^\1(N) \right) 
  \end{array}  
\end{equation}
On the zero mode side, the  Chern character for the (integer) representation of spin
$\mathbf{t}$ is given by
\begin{equation}
  \label{eq:59}
     \ch_\mathbf{t}(F) = (2t+1) + \frac{t(t+1)(2t+1)}{6}p_1(N) + \cdots
\end{equation}
and the total anomaly (sum over $t$ from 0 to $n-1$) is
\begin{equation}
  \label{eq:60}
  I_\zm =\int_\brane\left(\frac{\pi}{12}n^2\,p_1^\1(T) - \frac{\pi}{6} n^2(n^2-1)\,p_1^\1(N)\right)
\end{equation}

We see that for arbitrary fermion representation the anomaly cancels for
a simple choice of $\so3_{\diag{}}$ representation for the zero modes.

\section{Monopole with Charge $N>1$}
\label{sec:big-charge}

Recall~\cite{hmm} that for $N$  M5 branes the cancellation of normal bundle
anomalies from the modified Chern-Simons term predicts 
that the anomaly on the brane  scales like $N^3$ for large $N$.
This is related to the simple fact that the Chern-Simons term in 11-d SUGRA is
cubic in the three-form gauge field. If the magnetic charge of the field is $N$ one
expects that the magnetic coupling equation will have the form $dF_k=N d\rho
\wedge e_k$ and naively its solution will be just $F_k=N F_k(1)$ (where
$F_k(1)$ is the  charge one solution), so that  the Chern-Simons evaluated on such a
field will be proportional to $N^3$. In \cite{hmm}  
this information was used to deduce some
properties of the correlators of the $(2,0)$ theory. In our case, however, the
structure of fermionic zero modes is known, specifically it is
known that  the number of zero modes grows
as $N$~\cite{Callias}. Thus it is clear that the anomaly computed in the
world-volume theory on the monostring will only scale as $N$, and not as $N^3$.

To explain this difference between the M5-brane and the monostring, we note
that for the charge $N>1$ there are \emph{no} spherically symmetric solutions
in our model~\cite{wg}. The background and therefore the theory are not
symmetric inside the monopole core.  As a result there is no $SO(3)_{\rm diag}$
symmetry of
the world-volume theory and no anomaly to be computed.

To understand this fact better one can use the following example.  We can
build an explicit mapping $\Phi$ of degree $N$ from $S^2_\infty \rightarrow
S^2_{\su2}$, such that the  topological charge $Q_{\mathrm{top}} = N$. One way to
construct  such a mapping is to use stereographic projection from
$S^2$ on the complex plane and consider a map $w=z^N$ in complex coordinates.
It is easy to see that this mapping is not spherically symmetric -- the
Jacobian of the map depends on angles and  $\Tr(\Phi\wedge d\Phi\wedge
d\Phi) \ne N e_2$ (although it is equal to $N$ after integration over the
sphere).  Therefore, the  field strength will not scale as $N e_2$ and the
Chern-Simons will not necessarily give an $N^3$ contribution.

We see that in our model, the  fundamental theory which lives in the core of the
monopole does not posses the symmetry with respect to normal bundle rotations
for $N>1$ and this fact is not obvious  from the effective theory in
the bulk.  This raises the question of whether the analysis of \cite{hmm}
should really be taken as evidence for $N^3$ scaling.

For $N$ M5-branes there are independent arguments which strongly suggest the
$N^3$ dependence.
First, as usual for extremal p-brane solutions the area of the horizon is
related to the charge (the number of the branes) $N$ and thus
the Bekenstein-Hawking entropy depends on $N$. It is easy to see that for the case
of $N$ M5 branes it scales like $N^3$~\cite{klebs}. Similar arguments,
involving absorptions cross-section of low energy gravitons in the background
of $N$ coincident M5-branes, were discussed in~\cite{gubser}.
This provides an indication that the $(2,0)$ theory has a total number of
degrees of freedom that scales like $N^3$.  This of course does not give a
direct proof that the normal bundle anomaly scales like $N^3$.  The theory for
$N>1$ is interacting and its structure is still unclear.

%
However, independent information also arises from the calculation of the Weyl
anomaly in the $(2,0)$ theory, which also scales as $N^3$~\cite{henns}, and
the fact that the Weyl anomaly and the $SO(5)$ anomaly are in the same
supersymmetry multiplet. 


\section{Comparison with M-theory and Open \\ Questions}
\label{sec:other-anomalies}

It is interesting to push the analogy between the M5-brane and the monostring
as far as possible.  We can think of our \su2 theory~(\ref{eq:1}) as an analog
of M-theory in $D=11$, and the effective $U(1)$ theory with
terms~(\ref{eq:40}) -- as an analog of eleven-dimensional supergravity
together with the higher-derivative correction $G_4\wedge
X_8^\0$~\cite{witten,dlm}. In the case of eleven-dimensional supergravity
information about zero modes living on one M5 brane comes from a symmetry
analysis, that is they can be viewed as Nambu-Goldstone zero modes 
\cite{chs,kapmich,swedes}.  To make the analogy even more transparent, we
consider the supersymmetric version of our problem and then comment on some
other aspects.

We can easily construct a supersymmetric version of both the \su2 and $U(1)$
theories. It is given by starting from $(1,0)$ supersymmetric Yang-Mills
theory in $D=6$ and reducing it on an $S^1$. This gives a supersymmetric
theory in $D=5$ with  adjoint fermions, an adjoint scalar, and gauge
fields. The potential for the scalar field vanishes, and we can then consider
monopole solutions in the BPS limit of vanishing potential.

It is clear from thinking about this from the $D=6$ point of view that the
monopole-string breaks half of the supersymmetry and should lead to a
world-sheet sigma-model with $(4,0)$ supersymmetry.  Repeating the analysis
analogous to that in \cite{chs,swedes}, we see that the $(4,0)$ theory on the
magnetic string consists of the following modes: 3 (non-chiral)
Nambu-Goldstone scalars from breaking of translational invariance; 4
Mayorana-Weyl fermions from broken SUSY; one dyon from broken large gauge
transformations.  We see that the field content of this two-dimensional $(4,0)$
theory closely  resembles  that of the six-dimensional theory on the world
volume of one M5 brane. In the latter case the zero modes consist of five
scalars, 4 Weyl fermions and one self-dual two-form.  To make the analogy
complete, we would need a self-dual one-form in two dimensions, that is, a chiral
dyon zero mode. The issue of whether the dyon mode is chiral is in fact
somewhat intricate.

First of all, from the classical analysis of the full \su2 model there is
\emph{no} chiral dyon (c.f.~(\ref{eq:74})) in the spectrum of zero modes of
the theory. The question arises as to whether the dyon zero mode might be
chiral in the quantum theory.  It is interesting to note that $(4,0)$ SUSY
itself does not itself require a chiral dyon.  Indeed, every scalar field can
be divided into left and right moving parts.  $(4,0)$ supersymmetry acts only
on the left-moving components (of both bosons and fermions), leaving the
right-movers invariant.  Therefore the presence of right-moving scalar degrees
of freedom is not controlled by supersymmetry in any way.

For the M5-brane an analysis of the \emph{classical} theory shows that
the two-form zero mode is chiral. Both chiralities appear as zero modes,
but because of the Chern-Simons term, only one of them is normalizable.

On the other hand, a chiral dyon can be found in our model as well.  It
arises as a Goldstone zero mode, precisely as it appeared in the case of the M5
brane. Indeed, let us consider the  effective $U(1)$ theory with added Chern-Simons
term~(\ref{eq:40}). Recall that it was obtained as a quantum correction from
integrating out massive fermions.  Note that this is in contrast to the $D=11$
case, where the  Chern-Simons term was present in the action at  tree level (it was
required by supersymmetry) and therefore played a role in the classical
analysis of the zero modes. We can formally repeat the analysis 
of~\cite{chs,kapmich,swedes} for our case with the Chern-Simons
corrections~(\ref{eq:40}) and find that there is a localized massless zero
mode coming from large gauge transformations and that only part with one
chirality is normalizable (has decreasing radial profile). For details we
refer to Appendix~\ref{sec:dyon}.

How should we interpret this result and its apparent discrepancy with the
classical analysis~(\ref{eq:74})? To begin with, let us stress once again
that it is \emph{necessary} to have a Chern-Simons (or some other parity-odd)
term in the action in order to obtain a chiral dyon zero mode. This term was
absent in the original \su2 theory (and therefore the analysis there gave a
dyon zero mode with both chiralities present) and appeared only in the quantum
effective action as a result of interaction between the gauge field and
massive fermions.

One of the arguments in favor of the chiral two-form for the M5-brane was
anomaly cancellation: chiral two-form gives an additional contribution to the
tangent bundle anomaly and without it cancellation would not be
possible~\cite{witten}.  However in our case the situation is precisely the
opposite.  Indeed, let us suppose that the dyon is chiral in the $U(1)$
effective theory. Then together with 4 Mayorana-Weyl fermions it produces the
following anomaly:
\begin{equation}
  \label{eq:61}
  I_\zm =  \dint_\brane\left(\dfrac\pi{4} p_1^\1(T\Sigma)-\dfrac\pi 2 p_1^\1(N)\right)
\end{equation}
This anomaly differs by  $\frac\pi{12} p_1^\1(T\Sigma)$  from anomaly
inflow~(\ref{eq:50}). 

Before discussing this further, let us emphasize precisely what was involved
in studying anomaly cancellation so far.  In
Sections~\ref{sec:setup}---\ref{sec:anom-canc} we made an attempt to compute
the $\log\det\Dsl$ -- fermionic determinant in the background of
monopole~\mon.  We split the integration over the fermions $\Psi$ into two
parts:
\begin{equation}
  \label{eq:62}
  \int \CD\,\Psi\, e^{i S[\CA,\Psi]}\equiv\int \CD\,\Psi_0\CD\,\Psi'\, e^{i
    S[\CA,\Psi]} =  \int \CD\Psi_0\, e^{i S_{eff}^{(5)}[\CA] + i S^{(2)}[\Psi_0]}
\end{equation}
Here by $\CD\Psi'$ we mean integration over \emph{all but zero modes}, and by
$\CD\Psi_0$ we mean integration over the set of Jackiw-Rebbi zero modes.
Action $S_{eff}^{(5)}[\CA]$ was computed under assumption that $\Psi'$ is the
full system of functions and that the background is topologically trivial.
Such assumption had its price: we saw that Chern-Simons terms ~(\ref{eq:40})
in $S_{eff}^{(5)}[\CA]$ was anomalous, even though the original
Lagrangian~(\ref{eq:1}) was anomaly-free.


The action $S^{(2)}$ is the action of Jackiw-Rebbi zero modes on the string
described by eq.~(\ref{eq:24}) or~(\ref{eq:26}), depending on the
representation.

We would like to stress once again that eq.~(\ref{eq:62}) is just an
approximation, widely used in analyzing different brane constructions
(c.f.~\cite{ch,witten,hmm,fhmm,jh-or}).  The aim of this paper
is in particular to check the consistency of such approximation and to show
how many subtleties are encountered on the way of realization of this idea.
Thus, in Section~\ref{sec:anom-canc} we explicitly analyzed all anomalies
coming from representing the fermionic determinant as the sum of these two
actions and demonstrated that anomalies indeed cancel. This is the reflection
of the fact that in five dimensions there exists a gauge-invariant
regularization of fermions and $\log\det\Dsl$ is non-anomalous.

Having said all that, let's return back to the discussion of the dyon anomaly.
In the manner similar to that of described above 
we would in principle need to perform integration in two stages -- integration
over the zero mode (dyon) and integration over the gauge fields in the bulk.
Such integration in the bulk could give rise (among other things) to
renormalization of coefficients in~(\ref{eq:40}) and as a result change of the
inflow~(\ref{eq:50}).  We claim, however, that this is not the case. Indeed,
renormalization of parity-odd terms can come from parity-odd interactions
only, and Chern-Simons terms itself is the only candidate. However,
a simple analysis
(Appendix~\ref{sec:cs-int}) shows that this is not the case.

The dyon anomaly computation (i.e. integration over the dyon zero mode) is
intrinsically different from the fermion one. In contrast to the former case
dyon (as ``zero mode'' of gauge field) cannot be clearly separated from the
rest of the fields in path integral, as part of the \su2 gauge field
still remains massless. In this sense it is not even clear whether the dyon
zero mode ever decouples from the bulk in a sense in which Jackiw-Rebbi zero
modes did. This might suggest, that one cannot even write a local counterterm
for the dyon anomaly cancellation. However, we are not addressing this issue
in the current paper.

Another important difference between our model and the M-theory situation is
the origin of the Chern-Simons term. In the present analysis the Abelian
Chern-Simons arises as the one-loop quantum correction, on the same footing as
the term containing first Pontriagin class in~(\ref{eq:40}). As it was
discussed above, this Chern-Simons term is the bulk part of the fermionic
determinant.  Because of this it is clear why it has exactly the correct
coefficient in front of it to cancel the anomaly of normal bundle, which came
from computation of the same determinant. In the M5-brane case, however, the
situation is different. The ``Gravitational'' term $G_4\wedge
X_8^\0$~\cite{witten,dlm} did arise as a quantum correction, while the
Chern-Simons term was the (required) part of the ``tree-level'' $D=11$ SUGRA
action and the coefficient on front of it was fixed by supersymmetry.  At
first sight it is not clear at all why this terms would ``know'' anything
about quantum effects -- anomalies.  Note, that this problem is directly
related to the problem of chiral dyon discussed above.  If the parity-odd
Chern-Simons term is present from the very beginning at tree-level, the
``dyon'' (two-form) zero modes is chiral and plays its role in the
cancellation of anomalies. If it appears only as a one-loop correction, then
the classical dyon is non-chiral and does not contribute to anomalies.

There exists a direct five-dimensional analog of $D=11$
supergravity~\cite{d5sugra}. $N=1$ $D=5$ SUSY has a gravitational multiplet,
with bosonic degrees of freedom consisting of the
graviton and a $U(1)$ vector field. The vector
field part of such a supergravity Lagrangian has a  Chern-Simons term $A\wedge
F\wedge F$ and higher-derivative term $F\wedge p_1^\0(R)$ as in~(\ref{eq:40})
but with different coefficients, dictated by supersymmetry.  This SUGRA has
a magnetic p-string solution.  Goldstone zero modes analysis of this solution
gives \emph{the same} $(4,0)$ theory on the string (3 scalars, 4 Mayorana-Weyl
fermions, 1 chiral scalar).  But in \emph{this} case anomaly of $(4,0)$ theory
is precisely canceled by anomaly inflow~\cite{fhmm,d5sugra}.

The situation with the origin of Chern-Simons term in M-theory and its role in
anomaly cancellation resembles Green-Schwarz anomaly cancellation~\cite{gsw}.
{}From the viewpoint of the present analysis it could be interpreted as one more
evidence of the existence of some fundamental theory which would explain this
phenomenon naturally, as for string theory.

\section{Conclusions}
\label{sec:Conc}

In this paper we have studied the following field theory: Yang-Mills coupled
to a scalar Higgs field and fermions in five dimensions. This theory possesses a
monopole solution, which looks like a magnetically charged string. We
considered this theory as an example, which reproduced many features of the M5
brane anomaly cancellation. For example, we showed that both gauge and
gravitational anomalies of the fermion zero modes are canceled by anomaly inflow,
coming from the Chern-Simons terms which arise from integrating out
massive fermions in the bulk. We showed that the Chern-Simons terms, 
defined on a
topologically non-trivial background, possess  additional non-trivial
variation under the normal bundle transformation and thus contribute to the
normal bundle anomaly inflow (both in our example and in supergravity).

This analysis allowed us to make some predictions regarding the structure of
the fermion zero modes of the 't Hooft-Polyakov background for arbitrary
fermion representation.

Apart from that we were able to see that effective theory ($U(1)$ gauge theory
in our example) does not allow to resolve some important features of the full
(\su2) theory. Symmetry of the solution in the effective theory (Dirac
monopole in our case) does not coincide with the symmetry of the non-Abelian
monopole solution and therefore naive anomaly analysis gives the wrong
answer.  This shows once again that in case of several M5 branes the
conclusions drawn from analyzing the anomaly inflow in the effective theory
should be taken with the grain of salt.

The part in which our example differed the most from the original five-brane
case was the question of the chiral bosonic zero modes, arising as the
Nambu-Goldstone boson of large gauge transformations. In our case we were not
able to show what cancels the anomaly coming from the chiral dyon on the
world-volume on the string. We did not address this issue in detail, but we
argued that it was related to the fact, that in our case Chern-Simons was
generated at the one-loop level via integrating out massive fermions, while in
the eleven dimensional supergravity it was present from the very beginning. We
leave this and related questions for future investigations.

\section{Acknowledgments}
\label{sec:Ack}

We would like to thank G.~Bonelli, G.~'t~Hooft, B.~Kulik and E.~Witten for
discussions. This work was supported in part by NSF Grant No. PHY-9901194.
A.B. acknowledges support of Danish Research Council.

\appendix
\setcounter{equation}{0}
\renewcommand{\theequation}{\Alph{section}\arabic{equation}}

\section{Gamma Matrices}
\label{sec:gamma-matrices}

Following Jackiw and Rebbi~\cite{Jackiw-Rebbi} we take the following basis of
gamma-matrices (as usual: $\beta = \g0;\quad \vec \alpha =
\g0\cdot\vec\gamma$): 
\begin{equation}
  \label{eq:63}
  \g0 =  -i \left ( 
    \begin{array}[c]{cc}
      0 & 1\\
      -1 & 0
    \end{array}
    \right )
\end{equation}
\begin{equation}
  \label{eq:64}
  \g{k} =  -i \left ( 
    \begin{array}[c]{cc}
      \sigma_k & 0\\
      0 & -\sigma_k
    \end{array}
    \right )
\end{equation}
where $\sigma$ are the usual Pauli matrices:

\begin{equation}
\label{eq:65}
\sigma^1 = \left (
  \begin{array}[c]{cc}
    0 & 1 \\
    1 & 0
  \end{array} \right ), \quad
\sigma^2 = \left (
  \begin{array}[c]{cc}
    0 & -i \\
    i & 0
  \end{array} \right ), \quad
\sigma^3 = \left (
  \begin{array}[c]{cc}
    1 & 0 \\
    0 & -1
  \end{array} \right )
\end{equation}
One can easily check that we are working in $\eta_{\mu\nu} =
\diag(1,-1,-1,-1)$ signature. Define also
\begin{equation}
  \label{eq:66}
  \g4 = \prod_{n=0}^3 \gamma^n = 
  -i \left ( \begin{array}[c]{cc}
      0 & 1\\
      1 & 0
    \end{array} \right )
\end{equation}
Which also obeys $(\g4)^2 = -1$ and $\{\g4,\,\g{k}\}
=\{\gamma_4,\,\gamma_0\} = 0$
Let us also introduce the matrix $\g{int}$:
\begin{equation}
  \label{eq:67}
  \g{int} \equiv \g0 \g4=\left ( \begin{array}[c]{cc}
      -1 & 0\\
       0 & 1
    \end{array} \right )
\end{equation}

\section{Non-renormalizability of the Coefficient of the Chern-Simons.}
\label{sec:cs-int}

In this Section we show that coefficient of Chern-Simons does not get
renormalized due to the Chern-Simons interaction at one-loop
order. The analysis is very
simple. Interaction term $A dA dA$ gives the following vertex:
\begin{equation}
  \label{eq:69}
  V(k^\1,k^\2,k^\3) =
  (2\pi)^5\delta(k^\1+k^\2+k^\3)\epsilon_{\mu\nu\lambda\rho\sigma}A_\mu(k^\1)
  k^\2_\nu A_\lambda(k^\2) k^\3_\rho A_\sigma(k^\3) 
\end{equation}
%
%
Renormalization of the Chern-Simons term comes from a triangle diagram
with the insertion of~(\ref{eq:69}) in each of three vertices. This
diagram is proportional to
\begin{equation}
  \label{eq:71}
  S_{triang}\sim k^\1_\tau A_\mu(k^\1)
  k^\2_\nu A_\lambda(k^\2) k^\3_\rho A_\sigma(k^\3) 
\end{equation}
So, we see that interaction due to the Chern-Simons term cannot produce
counter-terms proportional to~(\ref{eq:69}).

\section{Classical and Goldstone dyon}
\label{sec:dyon}

We once again consider the Lagrangian~(\ref{eq:1}). Our conventions for the field
strength and covariant derivatives are
\begin{equation}
  \label{eq:72}
  \arstretch{1.5}
  \begin{array}{lcl}
    G^a_{M N} &= &\partial_M A^a_N - \partial_N A^a_M +
    g\epsilon_{abc} A^b_M A^c_N\\
    (D_M\Phi)^a &=& \partial_M \Phi^a + g \epsilon_{abc} A^b_M \Phi^c\\
    (D_M\psi)_n &=& \partial_M \psi_n - ig T^a_{n m} A^a_M \psi_m\\
  \end{array}
\end{equation}
In the absence of fermions, the equations of motion are:
\begin{equation}
  \label{eq:73}
  \arstretch{1.5}
  \begin{array}{l}
    D^N G_{N M}  = [D_M\Phi,\,\Phi]\\
    D^2 \Phi = -2\Phi\, U'(g\Phi)
  \end{array}
\end{equation}
We consider an  ansatz for the zero mode in the following form:
\begin{equation}
\label{eq:74}
  \left\{
    \begin{array}{l}
      \delta_\alpha A_\mu^a = 0\\
      \delta_\alpha \Phi^a = 0\\
      \delta_\alpha A_i^a(x,y)=  \alpha(x)D_i(f(y) \Phi^a)
    \end{array}
    \right.
\end{equation}
(recall that $y = \sqrt{y_i y^i}$ is the radial coordinate in transversal
directions). Here $f(y)$ is a profile function which needs to be determined.
Substituting this ansatz into the action, we find that variation~(\ref{eq:74})
over the background~\mon\ is still a solution of~(\ref{eq:73}) if
\begin{equation}
  \label{eq:75}
    \left\{
    \begin{array}{l}
      \p^2_x \alpha = 0\\
      f \sim 1/y,\quad y\rightarrow\infty
    \end{array}
  \right.
\end{equation}
Thus, we see that zero mode (scalar $\alpha(x)$) enjoys the following
properties: (1) it behaves like massless scalar in $1+1$ dimensions; (2) it is
localized inside the core, i.e. its transversal profile is given by $D(f\Phi)$
and decays as $1/y$ at infinity.  One should note that this derivation will
not be true in BPS case, when $m_H=0$ and $\Phi$ falls of much slower. In that
case, however, the dyon ansatz is also given by~(\ref{eq:74}) with $f=1$ (see,
e.g.~\cite{Harvey-mon}). Note, that the dyon $\alpha(x)$ is non-chiral from
the two-dimensional point of view!

However, things change once we consider this zero mode in the effective $U(1)$
theory~(\ref{eq:36}). In this case a dyon ansatz similar to~(\ref{eq:74}) is
given by:
\begin{equation}
  \label{eq:76}
  \delta_\alpha A = \alpha(x) d f(y);\qquad
  \delta_\alpha F = d\alpha(x)\wedge d f(y)
\end{equation}
with $f(y)$ being some normalizable function in the transverse directions. We
want to repeat the analysis of~\cite{chs,kapmich,swedes}. Let us forget
about gravity for simplicity (i.e. put $p_1(R)=0$ in~(\ref{eq:36})) and try to
find the equations of motion for $\alpha(x)$, using the ansatz~(\ref{eq:76}). We
will be somewhat schematic through the end of this Section, concentrating only 
on the details and ignoring all numerical factors.

Consider the $U(1)$ action
\begin{equation}
  \label{eq:77}
  S=\int F\wedge * F + \frac{c}{3} A \wedge F\wedge F
\end{equation}
(where all the numerical factors are absorbed into the constant $c$. Equations 
of motion reads:
\begin{equation}
  \label{eq:78}
  d*F - c F\wedge F = 0
\end{equation}
Substituting in~(\ref{eq:78}) $F=2\pi e_2 + d\alpha(x)\wedge d f(y)$ we get
(note that in the case of trivial normal bundle connection $e_2$ reduces down
to the volume form on two sphere $\domega=\sinq\;d\theta\wedge d\varphi$):
\begin{equation}
  \label{eq:79}
  d*(d\alpha\wedge d f) - 2c d\alpha\wedge d f\wedge e_2
\end{equation}
\begin{equation}
  \label{eq:80}
  \begin{array}{rcl}
  d*(d\alpha\wedge d f) & = & c_1 d\Bigl((f'(y) y^2)\epsilon_{\mu\nu}\p_\mu \alpha
  \,dx^\nu\wedge \domega\Bigr) \\ 
  & = & c_1 \Bigl(\p^2\alpha\, \epsilon_{\mu\nu} dx^\mu\wedge dx^\nu +\\
  & + & (y^2f')' dy\wedge
  \epsilon_{\mu\nu}\p_\mu \alpha\, dx^\nu)\Bigr) \wedge \domega
\end{array}
\end{equation}
Prime denotes the differentiation with respect to $|y|$ (we are working in
spherical coordinates in the $y$ directions). Constant $c_1$ absorbed some
numerical factors and/or signs, coming from performing operation $*$ and
differentiation.  We are not following them, as it is not really important,
because we are not interested in the precise chirality of the dyon, we only
want to know whether the object is chiral.

Now consider second term in~(\ref{eq:79}):
\begin{equation}
  \label{eq:81}
  -2 c \p_\nu\alpha f'(y) dx^\nu\wedge dy\wedge \domega
\end{equation}
we see that for eq.~(\ref{eq:80}) to cancel~(\ref{eq:81}) one needs two
conditions. First, form proportional to $dx^\mu\wedge dx^\nu\wedge \domega$
should be equal to zero, which means to $\p_x^2\, \alpha = 0$. For the
second term in~(\ref{eq:80}) we get
\begin{equation}
  \label{eq:82}
    c_1 \epsilon_{\mu\nu}\p_\nu \alpha (y^2 f')' - 2 c \alpha_\mu f' = 0
\end{equation}
which can be solved in the following way:
\begin{eqnarray}
  \renewcommand{\arraystretch}{2}
  \epsilon_{\mu\nu}\p_\nu\alpha= \pm\p_\mu\alpha\label{eq:83}\\
  (y^2 f')' = \pm\dfrac{2c}{c_1} f'\label{eq:84}
\end{eqnarray}
Solution of eq.~(\ref{eq:84}) is given by:
\begin{equation}
  \label{eq:85}
  f(y) = f_0 \left(1 - \exp\left(\mp \frac{2 c_1}{c y}\right)\right)
\end{equation}
Only for one choice of sign in~(\ref{eq:83}) ansatz~(\ref{eq:76}) describes
normalizable zero mode (chiral or anti-chiral, depending on the initial sign
of $c_1$). Resulting dyon falls off as $1/y$ for $y\rightarrow\infty$.

\end{document}